\documentclass[aps,prb,twocolumn,groupedaddress,showpacs,showkeys,citeautoscript,amsfonts,amsmath,amssymb,10pt,flushbottom]{revtex4-1}

\usepackage{graphicx}
\usepackage{nicefrac}
\usepackage[latin1]{inputenc}
\usepackage{amsmath}
\usepackage{mathtools}
 \usepackage[dvips]{color}

 \newif\ifshowcolor

 \ifshowcolor
 \newcommand{\bl}[1]{\textcolor{blue}{#1}}
 
 \else
 \newcommand{\bl}[1]{#1}
 
 \fi

\begin{document}
\newcommand{\bskipdm}{\mskip -3.8\thickmuskip}
\newcommand{\bskiptm}{\mskip -3.0\thickmuskip}
\newcommand{\bskipsm}{\mskip -3.1\thickmuskip}
\newcommand{\bskipssm}{\mskip -3.1\thickmuskip}
\newcommand{\fskipdm}{\mskip 3.8\thickmuskip}
\newcommand{\fskiptm}{\mskip 3.0\thickmuskip}
\newcommand{\fskipsm}{\mskip 3.1\thickmuskip}
\newcommand{\fskipssm}{\mskip 3.1\thickmuskip}
\newcommand{\pint}{\mathop{\mathchoice{-\bskipdm\int}{-\bskiptm\int}{-\bskipsm\int}{-\bskipssm\int}}}
\newcommand{\ds}{\displaystyle}
\newcommand{\D}{\mathrm{d}}
\newcommand{\I}{\mathrm{i}}
\newcommand{\EXP}[1]{\mathrm{e}^{#1}}    

\newcommand{\tpp}{\hat{t}}
\newcommand{\rpp}{\hat{r}}
\newcommand{\zpp}{\hat{z}}
\newcommand{\ppp}{\hat{\varphi}}
\newcommand{\tbb}{\bar{t}}
\newcommand{\rbb}{\bar{r}}
\newcommand{\zbb}{\bar{z}}
\newcommand{\pbb}{\bar{\varphi}}
\newcommand{\lbb}{\bar{\ell}}
\newcommand{\ttt}{\tilde{t}}
\newcommand{\rtt}{\tilde{r}}
\newcommand{\ztt}{\tilde{z}}
\newcommand{\ptt}{\tilde{\varphi}}
\newcommand{\abs}[1]{\left\lvert #1 \right\rvert}
\newcommand{\tvv}{\check{t}}
\newcommand{\xvv}{\check{x}}
\newcommand{\omvv}{\check{\omega}}
\newcommand{\kvv}{\check{k}}
\newcommand{\vvv}{\check{v}}
\newcommand{\lvv}{\check{\ell}}

\title{Classroom reconstruction of the Schwarzschild metric}

\author{Klaus Kassner}
\affiliation{Institut für Theoretische Physik, \\
  Otto-von-Guericke-Universität Magdeburg, Germany }
\email{Klaus.Kassner@ovgu.de}

\begin{abstract}
  A promising way to introduce general relativity in the classroom is
  to study the physical implications of certain given metrics, such as
  the Schwarzschild one. This involves lower mathematical expenditure
  than an approach focusing on differential geometry in its full glory
  and permits to emphasize physical aspects before attacking the field
  equations. Even so, in terms of motivation, lacking justification of
  the metric employed may pose an obstacle. The paper discusses how to
  establish the weak-field limit of the Schwarzschild metric with a
  minimum of relatively simple physical assumptions, avoiding the
  field equations but admitting the determination of a single
  parameter from experiment.  An attractive experimental candidate is
  the measurement of the perihelion precession of Mercury, because the
  result was already known before the completion of general
  relativity. It is shown how to determine the temporal and radial
  coefficients of the Schwarzschild metric to sufficiently high
  accuracy to obtain quantitative predictions for all the remaining
  classical tests of general relativity.
\end{abstract}

\date{11 September 2015}

\maketitle

\allowdisplaybreaks

\section{Introduction}

Conceptually speaking, general relativity (GR) is not a particularly
difficult theory.  From the viewpoint of physics education, all the
conceptual impositions of the relativity theories arguably arise with
special relativity (SR) already. It is in courses of SR
that students will be exposed to the relativity of simultaneity and a
variety of paradoxes, having to do with time dilation and differential
aging,\cite{debs96,cranor99} length contraction and the pole-barn
paradox,\cite{vanderweele07} bridges that may or may not collapse
under relativistic trains,\cite{fayngold02} relativistic lever
experiments appearing to violate angular momentum
conservation,\cite{butler70} Bell's spaceship
paradox,\cite{dewan59,bell76} Ehrenfest's
paradox\cite{ehrenfest09,gron81a,kassner12} and the appearance of
non-Euclidean geometry in accelerating systems such as a rotating
disk.\cite{berenda42,gron75,kassner12} Those students will probably
not be overly shocked by the additional complication of
\emph{spacetime curvature} in GR.

And this is essentially the only conceptual complication. Some things
even get simpler with certain standard examples of GR
systems. In SR, we have the bewildering phenomenon of
mutual time dilation, utterly incomprehensible without an
understanding of the non-absoluteness of simultaneity. When comparing
coordinate stationary observers (CSOs) in a static metric, there
usually also is time dilation, but different observers agree on whose
clocks are runnig faster, a situation that is not especially difficult
to visualize.

What makes GR difficult, is the mathematical overhead.
SR can be taught with very little calculus, whereas in
GR, differential geometry is essential. The field
equations of GR are intrinsically nonlinear, so their
solution is, even in the simplest cases, nontrivial. The Riemann
curvature tensor has 20 independent components.

Given the conceptual simplicity and the mathematical complexity of GR, it
is natural to ask whether it is possible to find a simpler approach to
certain fundamental aspects of the theory, to make it more accessible to
students in the transition from special to general relativity. A
full-fledged course in GR will have to deal with the
field equations eventually, but the entry point into the theory might
be based on much simpler considerations. An SR course
giving a glimpse at GR near its end may benefit from
avoiding the field equations altogether.

It is a substantiated view that exploring the con\-sequences of a
particular metric (normally the Schwarz\-schild one) leads to an
accessible ``physics first'' approach to introducing
GR\cite{hartle06,christensen12}. Unfortunately, the metric will arise
out of the blue in such a strategy.  Therefore, it is legitimate to
inquire whether we can do better and obtain nontrivial metrics from
simple arguments, without going all the way to the field equations.

Such an idea was implicit in the so-called Lenz-Schiff argument,
apparently never published by Lenz but presented in Sommerfeld's
textbook\cite{sommerfeld52} and used by Schiff\cite{schiff60} to argue
that light deflection by the sun is quantitatively describable without
the field equations. This would put it on a par with the gravitational
redshift of spectral lines in the field of a weakly gravitating object
such as our sun, known to be explicable by a combination of special
relativity with the Newtonian limit (NL), using Einstein's equivalence
principle (EP). The perihelion precession of Mercury would then remain
the only one of the three classical tests of GR that really probes the
field equations.

Schiff's paper was shown to be in error.\cite{sacks68,rindler68}
Nevertheless, recurrently\cite{rowlands97} and even
recently\cite{cuzinatto11} articles have been published that ``derive''
GR effects requiring spacetime curvature on the basis of the
fallacious Lenz-Schiff argument. Yet, detailed arguments had been given
\cite{schild60,gruber88} that a simple derivation of the Schwarzschild
metric, i.e., one avoiding knowledge that traditionally is gathered
from the field equations, is impossible.  

The appearance of a controversy may be deceiving. Advocates of the
Lenz-Schiff argument seem to be unaware of its deficiencies. In
contrast, anyone familiar with the foundations of GR will realize that
\emph{any} nonsingular metric (with Minkowskian signature) is locally
compatible with SR, due to the equivalence principle (stating that it
always possible to transform the metric to Minkowski form locally).
Therefore, the EP does not \emph{constrain} the metric. Without the
field equations or some equivalent, constraints on the metric arise
only from symmetry and the NL. Any corrections to the NL,
expressible in powers of the small quantity $GM/c^2 r$ for a
spherically symmetric situation,\footnote{$G$ is Newton's
  gravitational constant, $c$ the speed of light, $M$ the mass at the
  center of gravity and $r$ the radial coordinate at which the metric
  is considered.} must be missed by a Lenz-Schiff type approach as, in
fact, by any other approach based on local considerations within the
framework of SR.  Therefore, no more than a weak-field approximation
to the metric can be gained.  This does not yet exclude Schiff's
result, essentially referring to the first-order term in powers of
$GM/c^2 r$ of the radial metric coefficient $g_{rr}$.  However, the
aforementioned analyses\cite{schild60,gruber88} show that the
Newtonian limit gives us only $g_{rr}=1$.  Thereafter, any claims to
deriving the Schwarzschild metric or even a post-Newtonian
approximation to it on the basis of just symmetry, SR, the EP, and the
NL, are recognizably erroneous.

To be precise, this does \emph{not} mean that one cannot do without
the \emph{field equations}.

Essentially, either the field equations or their generating action
including its Einstein-Hilbert part are a set of \emph{postulates} within GR.
In most theories based on postulates or axioms, the axioms are
\emph{not unique}.  In set theory, for example, the axiom of choice,
Zorn's Lemma and the well-ordering theorem are all
interchangeable.\footnote{To the uninitiated, these are qualitatively
  different axioms with very different levels of plausibility.}  It is
sufficient to postulate one of them. The other two are then derivable.
In thermodynamics, there are various different formulations of the
second law, which is a postulate of the theory. It is sufficient to
take one of them, then the others can be derived as
theorems.\footnote{Here, we are so accustomed to their equivalence
  that we do not even see them as different postulates, which formally
  they are.}

If consideration is restricted to the static spherically symmetric
case, it may be possible to use a simpler postulate (or two) than the
one leading to the field equations to derive a metric. Due to the
restriction to spherical geometry, there is no need that the postulate
be powerful enough to replace the field equations altogether. It is
sufficient, if it can replace them in spherically symmetric
situations.\footnote{Hence, the postulate(s) should be derivable from the
  field equations but need not imply their general form.} This kind of
approach is not only logically possible, it has even been discussed
favorably by Sacks and Ball\cite{sacks68} with regard to Tangherlini's
postulational approach to the Schwarzschild
metric.\cite{tangherlini62} Unfortunately, Rindler later showed one of
the two Tangherlini postulates to be unconvincing.\cite{rindler69} But
clearly, Tangherlini's approach is not subject to the criticism (nor
the impossibility proof) offered by Gruber \emph{et
  al.}\cite{gruber88}

Since both postulates from Ref.~\onlinecite{tangherlini62} cannot be
used here, the exact Schwarzschild metric will not be obtained.
However, I use one argument beyond the aforementioned ingredients
(symmetry, SR, EP, NL) to restrict the form of the metric. This will
reduce to the more convinicing one of Tangherlini's two postulates,
and it will be better justified than his statement. On the other hand,
additional information will be needed to obtain a truly post-Newtonian
approximation, and this can be taken from \emph{experiment}.
Amusingly, all of this information was available in 1911, when
Einstein published a calculation of light deflection by the
sun,\cite{einstein11} reproducing von Soldner's century-old
result\cite{vonsoldner04} and thus missing the correct prediction by a
factor of 2.

The general outline of the paper is as follows. In
Sec.~\ref{sec:rindler}, the metric describing the closest
approximation to a uniform gravitational field that is
relativistically possible, also known as the Rindler metric, is
derived. This simple problem, rigorously solvable within special
relativity, serves to expose the interplay of symmetry arguments and
thought experiments allowing us to obtain metric coefficients, an
approach that then may be applied to more complex situations. It also
demonstrates a little-appreciated property of the Rindler metric
justifying its interpretation as describing a uniform gravitational
field.  Section
\ref{sec:schwarzschild} is devoted to an introductory attempt at
constructing the metric of a spherical mass distribution using
symmetry, the EP and the NL.  At first sight, this
approach succeeds in obtaining the exact Schwarzschild metric. Its
deficiencies become visible on analysis of the order of approximation
achieved. A plausible and simple physical assumption 
turns out to partially cure the problem.  In
Sec.~\ref{sec:perihelion_prec}, it will be shown how information from
a true experiment,\footnote{In the present context, by a true
  experiment as opposed to a thought one I mean an experiment the
  outcome of which cannot be predicted quantitatively based on
  theories known before GR.}  \textit{viz.}~measurement of the perihelion
precession of Mercury, may then be used to  resolve the issue and to obtain
the weak-field limit of the Schwarzschild metric with sufficient
accuracy to quantitatively predict light deflection by the sun,
as demonstrated in Sec.~\ref{sec:light_bending}, and the Shapiro
delay.\cite{shapiro64} That section  also discusses
how the aforementioned factor of 2 can be found without actually
performing the full calculation. Finally, some conclusions are given
in Sec.~\ref{sec:conclusions}.

Most of the material is presented as if we did not know GR yet, but
occasionally this stratagem is dropped to address teachers directly,
who are assumed to be more knowledgeable. We shall assume general
acquaintance with SR and the use of different coordinate
representations of the Minkowski metric.

\section{Derivation of the metric of a ``uniform'' gravitational field}
\label{sec:rindler}

Our first aim is to deal with the inertial field inside an accelerating
object, say a big spacecraft. This situation may be completely
described within SR. At each point, the acceleration is to be
constant in time, i.e., each observer feels a constant proper
acceleration. In Newtonian physics, we would get something resembling
a uniform gravitational field, if all observers had the same
acceleration. In relativistic physics, we know that if observers
arranged along the direction of acceleration had the same proper
acceleration, Bell's spaceship paradox\cite{dewan59,bell76} would
apply -- they would find each other moving apart. Rather, we are
interested in a situation that is considered static by all observers.
This is Born rigid motion \cite{born09}, in which the proper distances
between our aligned observers remain constant, which means that from
the vantage point of an inertial system leading observers must
accelerate more slowly than trailing ones, so their distance shrinks
precisely according to the appropriate Lorentz factor.

Let us introduce an inertial frame $\Sigma$ with time $T$ and
cartesian coordinates $X$, $Y$, $Z$ and have its $X$ axis oriented
parallel to the direction of acceleration. Consider first a single
(point-like) observer $O$.  Since his proper acceleration is constant,
he will feel a constant accelerating force $f_0$, which is also the
force, by which a momentarily comoving inertial observer 
(in frame $\Sigma'$) 
will perceive $O$ to be accelerated. Moreover, the relativistic
transformation law for forces parallel to the vector of relative
motion between inertial systems tells us that the force $F_0$, by
which $O$ is accelerated in $\Sigma$ is the same as in  $\Sigma'$: $F_0=f_0$. Then the
equation of motion for $O$'s trajectory in  $\Sigma$ reads
\begin{align}
\frac{\D}{\D T} m \gamma(V) V = f_0\>,
\end{align}
where $V(T)=\D X/\D T$ and $\gamma(V)=1/\sqrt{1-V^2/c^2}$. If we set
the time $T$ equal to zero at the moment when $V=0$, this is solved
by
\begin{align}
\gamma(V) V = a T\>,
\label{eq:gamma_V}
\end{align}
where $a=f_0/m$ is the proper acceleration of $O$. Solving for $V$, we have
\begin{align}
  V= \frac{aT}{\sqrt{1+\left(\frac{aT}{c}\right)^2}}\quad \Rightarrow
  \quad \gamma(V) = \sqrt{1+\left(\frac{aT}{c}\right)^2}
\label{eq:velocity}
\end{align}
and this can be integrated once more to obtain
\begin{align}
X =  \frac{c^2}{a}\left(\sqrt{1+\left(\frac{aT}{c}\right)^2}-1\right) + X_0\>.
\label{eq:hyperbolic_traj}
\end{align}
The trajectory $X(T)$ is a hyperbola rather than the parabola known
from Newtonian physics, hence the notion of hyperbolic motion.  Note
that by taking the time derivative in \eqref{eq:velocity}, we end up
with the standard relationship for the transformation of longitudinal
acceleration:
 $ A \equiv {\D^2 X}/{\D T^2} = {a}/{\gamma(V)^3}\>.$
Next, consider two observers $O_1$ and $O_2$, starting at $X_{01}$ and
$X_{02}$ with proper accelerations $a_1$ and $a_2$, so their
trajectories are given by
\begin{align}
X_1 &=  \frac{c^2}{a_1}\left(\sqrt{1+\left(\frac{a_1 T}{c}\right)^2}-1\right) + X_{01}\>,
\nonumber\\
X_2 &=  \frac{c^2}{a_2}\left(\sqrt{1+\left(\frac{a_2 T}{c}\right)^2}-1\right) + X_{02}\>,
\label{eq:hyperbolic_two}
\end{align}
and require the distance between them to remain constant in the frame
of the first.\footnote{The symmetry properties of the resulting
  formulas ensure that the distance is then constant in the frame of
  $O_2$ as well.} The Lorentz transformations from $\Sigma$ to an
inertial frame momentarily comoving with $O_1$ read
\begin{align}
  x &= \gamma \left(X-X_1 - V (T-T_1)\right)\>, \quad \gamma =
  \sqrt{1+\left(\frac{a_1 T_1}{c}\right)^2}\>,
    \nonumber\\
    t & = \gamma \left(T-T_1 -\frac{V}{c^2} (X-X_1)\right)
\end{align}
and they transform the point $(T_1,X_1)$ to the origin of the comoving
inertial observer, in whose frame the distance $D$ between $O_1$ and $O_2$
at this moment is obtained by setting $X=X_2$ and choosing $T=T_2$ so
that $t=0$. This gives 
\begin{align}
T_2&=T_1+ \frac{V}{c^2} (X_2-X_1)\>,
\nonumber\\
D&=x_2 = (X_2-X_1)/\gamma \>\>\Rightarrow\>\> T_2 = T_1 
+ \frac{\gamma(V) V}{c^2} D\>.
\label{eq:T_rel_and_D}
\end{align}
Requiring in addition that the velocity of $O_2$ with respect to $O_1$
be zero, we find $\left.\frac{\D X}{\D T}\right\vert_{X=X_2} = V$
(obviously). This implies $a_1 T_1 = a_2 T_2$, which together with
\eqref{eq:gamma_V} and \eqref{eq:T_rel_and_D} leads to $a_1 T_1/a_2 =
T_1 \left(1+ a_1 D/c^2\right)$. We then obtain the important
relationship
\begin{align}
D = \frac{c^2}{a_2} - \frac{c^2}{a_1} = X_{02}-X_{01}\>.
\end{align}
Thus, by choice of the origin of $\Sigma$, we may  achieve $X_{0i} =
c^2/a_i$ for observer $O_i$ where originally $i=1,2$, but evidently,
this can be extended to an arbitrary number of observers. Equation
\eqref{eq:hyperbolic_traj} for the trajectory of an observer starting
from $X_0=x$ then simplifies to 
\begin{align}
X = \sqrt{x^2 + c^2 T^2}\>.
\label{eq:hyperbol_motion}
\end{align}
If we fill a half-space with observers labeled by their positive
initial coordinate $x$ and have them move according to
\eqref{eq:hyperbol_motion} with their $Y$ and $Z$ coordinates
unchanged, the ensemble will perform Born rigid motion.

To obtain the metric describing the common rest frame of
these observers, we note that translational symmetry in the $y$ and
$z$ directions as well as the requirement of time independence of the
metric imply the following general form of the spacetime line
element  \footnote{In fact, there could be a term of the form $H(\tilde
   x) \,\D t \D \tilde x$, but this can be transformed away by a
   resynchronization transformation of time, as is briefly discussed for the
   spherically symmetric metric in Sec.~\protect\ref{sec:schwarzschild}.}
\begin{align} 
  \D s^2 &= -F(\tilde x) \,c^2 \D t^2 + G(\tilde x) \, \D \tilde x^2 +
  \D y^2 + \D z^2\>.
\label{eq:parallel_metric_1}
\end{align}
Here, we have temporarily garnished one coordinate with a tilde,
because for $G(\tilde x)\ne 1$, the proper length element $\D\ell$ of this
metric ($\D\ell^2= \D s^2\vert_{\D t=0}$, due to time-orthogonality) does not
have its standard form, whereas our relationship for the proper
acceleration derived above was formulated in terms of the proper
distance. The simple coordinate transformation
\begin{align} 
x(\tilde x) = \!\int^{\tilde x} \sqrt{G(u)} \,\D u
\end{align} 
turns \eqref{eq:parallel_metric_1} into
\begin{align} 
  \D s^2 & = -f(x) \,c^2 \D t^2 + \D x^2 + \D y^2 + \D z^2\>,
\label{eq:parallel_metric_2}
\end{align}
so all that remains to be determined is the function $f(x)$. 

The proper time of a CSO, i.e., an observer satisfying $\D x=\D y = \D
z=0$, is given by ($\D s^2 = -c^2 \D\tau^2$)
\begin{align}
\D \tau = \sqrt{f(x)}\, \D t\>,
\end{align}
so observers at positions $x_1$ and $x_2$ will find their standard
clocks be subject to time dilation according to
\begin{align}
  \frac{\D \tau_2}{\D \tau_1} =\sqrt{\frac{f(x_2)}{f(x_1)}}\>.
\label{eq:time_dil_rindler_cso}
\end{align}
To determine the time dilation factor, imagine that $O_1$ sends an
electromagnetic signal having the frequency $\nu_1$ to the very close
$O_2$, who will receive it at frequency $\nu_2$. A good inertial
frame to discuss this in is the frame momentarily comoving with
$O_1$ at the emission event. During the short time interval $\Delta
\tau = (x_2-x_1)/c$ taken by the light, $O_2$ will have picked up the
small velocity $v=a(x_2)\Delta \tau$ in this inertial frame. If
$x_2>x_1$, $O_2$ is moving away from the emission event (since both
the signal and $O_2$ are moving towards increasing $x$), so the
frequency $\nu_2$ on reception of the signal will be lower than
$\nu_1$, due to the Doppler effect. The relativistic Doppler shift is given by
\begin{align}
  \frac{\nu_1}{\nu_2} = \sqrt{\frac{c+v}{c-v}} \approx 1 + \frac vc
  \approx 1 + \frac{a(x_2)\Delta \tau}{c} \approx 1 + \frac{a(x_2)\Delta x}{c^2}\>.
\label{eq:doppler_rindler}
\end{align}
We introduce an acceleration potential via $\D \Phi = a(x)\, \D x$.
Then \eqref{eq:time_dil_rindler_cso} and \eqref{eq:doppler_rindler}
imply ($\D \tau_2/\D \tau_1 = \nu_1/\nu_2$), for sufficiently small $x_2-x_1$:
\begin{align}
\sqrt{\frac{f(x_2)}{f(x_1)}} =  1 + \frac{\Phi(x_2)-{\Phi(x_1)}}{c^2}\>,
\end{align}
which can be easily converted into a differential equation for $f(x)$,
\begin{align}
  \sqrt{\frac{f(x+\D x)-f(x)}{f(x)}+1} &= 1+\frac12 \frac{f'(x)}{f(x)}
  \,\D x 
 \nonumber\\
&= 1+\frac{1}{c^2} \Phi'(x) \,\D x \>,
  \nonumber\\
  \frac12 \frac{f'(x)}{f(x)} &= \frac{1}{c^2} \Phi'(x)\>,
\end{align}
and this is solved by
\begin{align}
f(x) = \EXP{2\Phi/c^2}\>,
\label{eq:f_strong_field_R}
\end{align}
where the integration constant can be chosen by fixing an additive
constant implicit in the definition of $\Phi(x)$. Since we know the
position dependence of the proper acceleration $a(x)$, it is
straightforward to obtain the potential:
\begin{align}
\frac{\D \Phi}{\D x} = a(x) = \frac{c^2}{x}\quad\Rightarrow\quad \Phi(x) = c^2 \ln x + \text{const.}
\end{align}
This gives
\begin{align}
f(x) = \frac{g^2 x^2}{c^4}\>,
\end{align}
where $g$ is the proper acceleration at the position $x_0$, where
$f(x_0)=1$.

Even though we have now successfully derived the Rindler metric
\begin{align}
  \D s^2 = -\frac{g^2 x^2}{c^4} \,c^2 \D t^2 + \, \D x^2 + \D y^2 + \D
  z^2\>,
  \label{eq:rindler}
\end{align}
it is useful to look at another thought experiment. Consider an
observer at $x_0$ who slowly lowers some small mass $m$, hanging from
an inextensible massless tether, towards smaller $x$ values. 

Of course, in relativity, there are no \emph{truly} inextensible
bodies, because they would allow infinitely fast signaling (pull at
one end to immediately transfer a message to the other).  However, we
do not need more than approximate inextensibility.  An
  inextensible tether is simply one with a very large Young's modulus
  $Y$. The larger we make $Y$, the better the approximation.  The kind
  of inextensibility we want here may be expressed in terms of the
  covariant formulation of Hooke's law:\cite{gron81} All length
  changes required by relativistic kinematics are allowed, but local
  proper length increments of the tether do not change under a
  force.\footnote{Since real tethers always violate this condition
      to some extent, we might contemplate various contraptions to
      realize it to arbitrary precision. One possibility would be to
      make the tether a chain consisting of very short links and to
      monitor its local length changes by an appropriate device
      attached to it. Each time a local length unit is stretched by
      the extension of one link, the device removes one link and joins
      the separated parts of the chain; each time a local lengh unit
      is compressed by the length of a link, a link is inserted by the
      device.}  The only reason for this requirement is to avoid the
  consideration of elastic or plastic effects. Since we use our
tether for quasistatic transport only, fast signaling will not
occur.\footnote{A similar statement applies to the attribute
    \emph{massless}, which is also not realizable exactly.}

What is important in the following is that if a piece $\D \ell$ of the
tether is threaded down at its upper end, the lower end will move down
by the same amount $\D \ell$ in terms of its local proper length. The
question we ask then is: what is the force needed to hold the mass
at position $x$? At the beginning of the process, i.e., at $x_0$, we
clearly expect the force to be $F=-m g$, but as the mass is lowered,
it will experience different local proper accelerations. A way to
calculate the force is to invoke energy conservation. On being
lowered, the mass is doing work, so we should have
\begin{align}
  F = -\frac{\D E(x)}{\D \ell} = -\frac{\D E(x)}{\D x}\>,
\end{align}
where $E(x)$ is its energy at position $x$, as judged by the observer
at $x_0$. Now locally, the mass always has energy $m c^2$, as it does
not acquire kinetic energy -- the experiment is performed
quasistatically.  But the observer at $x_0$ will not assign this local
value to energy, because to him everything at $x$ happens at a slower
rate due to time dilation. This reduces the energy of photons by the
time dilation factor. Clearly, all other energies must be affected the
same way, otherwise no consistent physical description would be
possible.  To see this in more detail, imagine that the energy of
massive particles is reduced by a factor $\beta<1$, that of photons only
by a factor $\alpha>\beta$. Suppose the local observer has an electron
and a positron annihilate to produce two photons, the energy of which
locally is given by the sum of the particle energies: $2 h\nu =
E_{e}+E_{e^+}$. For the distant observer, energy conservation would be
violated, because the energy of the two photons would be $2 \alpha
h\nu$, that of the particles $\beta(E_{e}+E_{e^+})$, and $2 \alpha
h\nu >\beta(E_{e}+E_{e^+})$.  By an appropriate procedure, with the
lower observer sending photons to the upper one, who converts them
into particles that he sends down, where they are converted into
photons again, a perpetuum mobile (of the first kind!) could be
built.\footnote{\bl{To make sure that we never end up with photons
    whose energy is insufficient to create an electron-positron pair,
    the process could be started with such a pair at rest near the
    upper observer, so the initial energy would be $2 E_0$ with
    $E_0=E_{e}(v=0)=mc^2$. After the electron and positron are sent
    down to the lower observer, they will have acquired kinetic
    energy, so that their total energy remains $2 E_0$ (the rest
    energy now being $2 \beta E_0$) from the point of view of the
    upper observer. Conversion to photons will transform the
    \emph{total} energy into $2 \alpha E_0/\beta$ and the photons
    arriving back at the upper observer will then have a frequency
    $\nu = \alpha \nu_0/\beta$, where $h \nu_0 = E_0$. Now either the
    energy $2(\alpha/\beta-1) E_0$ can be extracted and the remainder
    used to create an electron-positron pair at rest, to be sent down
    and starting an identical repetition of the cycle.  Or else an
    electron-positron pair is created directly with some kinetic
    energy.  Repetition of the cycle will then gain another factor
    $\alpha/\beta$ in energy and energy extraction may be delayed
    until a sufficiently large chunk of energy is available to be
    immediately useful.}}

A perpetuum mobile could of course also be constructed, if
   $\alpha<\beta$, using the reversed sequence of
  processes.

From these considerations we conclude
\begin{align}
 E(x) &= \sqrt{\frac{f(x)}{f(x_0)}}\, m c^2 = \sqrt{f(x)}\, m c^2\>,
\label{eq:energy_of_x}
\end{align}
(because $f(x_0)=1$), hence
\begin{align}
F(x) =  -m c^2 \frac{f'(x)}{2\sqrt{f(x)}} \>,
\label{eq:force_metric_rindler}
\end{align}
which evaluates to $F = - m g$. Therefore, the force exerted by a mass
hanging from a tether is constant for a given observer,\footnote{It is
  different for observers at different $x_i$ values, as $f(x_0)$ in
  \eqref{eq:energy_of_x} gets replaced by $f(x_i)$. Hence, $F= -m
  a(x_i)$ for observer $O_i$.} no matter by how much it is lowered in
the ``inertial field'', a fact that has been noted by Gr{\o}n
before.\cite{gron77} This is the meaning of ``uniform'' when we are
talking about the \emph{uniform gravitational field} -- homogeneity of
the force on a particle or an object in a fixed observer's frame rather
than homogeneity of acceleration (i.e., homogeneity of the force per
unit mass). A detailed discussion of the issue of uniformity of fields
in GR is given in Ref.~\onlinecite{munoz10}.

Note that we could have \emph{derived} the metric by \emph{requiring}
the tether force to be constant and using
\eqref{eq:force_metric_rindler}. This derivation would be less
rigorous than the one actually given but it would be physically well
motivated and the result would be valid.  

The Rindler metric is related to the Minkowski metric by a coordinate
transformation
\begin{align}
c T  = x \sinh\frac gc t\>, \quad X =  x \cosh\frac gc t\>, 
\quad
Y = y \>, \quad Z = z \>,
\end{align}
so it describes a flat spacetime, the curvature of which necessarily
vanishes. In modern parlance, gravity is often identified with the
curvature of spacetime, but Einstein's view rather was that inertial
and gravitational fields are identical in nature. Moreover, while the
equivalence principle declares this identity only locally for
inhomogeneous gravitational fields, it is not inconceivable that a
mass distribution (homogeneous in $y$ and $z$) could be constructed
theoretically that would produce the metric \eqref{eq:rindler} in the
vacuum delimiting it above some $x$ value. Would we then refuse to
call the corresponding attractive field gravitational just because
spacetime happens not to be curved?

\section{The metric outside a spherical mass distribution}
\label{sec:schwarzschild}

Next, we would like to extend the ideas developed so far to a
nontrivial gravitational field, one that cannot be obtained by a
simple coordinate transformation from the Minkowski metric.

A leading theme in general relativity is to explain gravity in
  terms of spacetime geometry; in particular, the notion of
  \emph{curvature} of spacetime becomes important. A basic object in
  describing spacetime geometry is the metric. There are more complex
  objects, derivable from the metric (such as the Riemann curvature
  tensor), that may be used to decide whether the metric describes a
  flat or a curved spacetime. None of these objects will be needed
  here. It may be sufficient to say that if we write done some metric
  randomly, it is much more likely that it will describe a curved
  spacetime than not. A condition for flatness is that there exists a
  global coordinate transformation that takes our metric to Minkowski
  form. Since this is true only when certain integrability conditions
  are met (that are expressible via the Riemann tensor), most metrics
  that we may care to write down will comprise curvature and this is
  related to the presence of gravitation according to GR.

One of the simplest gravitating systems is a time-inde\-pendent
spherically symmetric mass distribution. \bl{We expect it to be}
describable by a static spherically symmetric metric. \bl{As we shall
  convince ourselves presently,} the line element may \bl{then,
  without loss of generality,} be written as
\begin{align}
\D s^2 &= -f(r) \,c^2 \D t^2 + g(r) \, \D r^2 
 + r^2 \left(\D \vartheta^2 + \sin^2 \vartheta \,\D \varphi^2\right)\>.
\label{eq:spherical_static}
\end{align}
Here, $\vartheta$ and $\varphi$ are the usual angular coordinates
which, due to spherical symmetry, may only appear in the combination
$\D\Omega^2= \D \vartheta^2 + \sin^2 \vartheta \,\D \varphi^2$ but not
in any of the coefficient functions. Because the metric is assumed
time independent, none of the coefficients may depend on $t$.
Therefore, all of them may be functions of the radial coordinate $r$
only. The pre\-factor of $\D\Omega^2$ might contain some additional
function of $r$, which we can however get rid of by redefining $r$ so
that the surface of any sphere about the coordinate center, given by
$r=\text{const.}$, becomes $4\pi r^2$. Finally, a term of the form
$h(r)\, \D t \,\D r$ would be allowed by symmetry, but can be
re\-moved by a coordinate transformation $ t \to t + w(r)$.\footnote{
  The expression $-f(r) \,c^2 \D \tilde t^{\,2} + h(r)\, \D \tilde t \,\D
  r+ \tilde g(r) \, \D r^2$ changes, on setting $\D \tilde t= \D
  t+w'(r) \D r$, into  $-f(r) \,c^2 \D t^2 + \bigl(h(r)-2 f(r)\, c^2
    w'(r)\bigr)\, \D t \,\D r+ \bigl(\tilde g(r)- f(r) \,c^2\, w'(r)^2
   +h(r) w'(r)\bigr)\D r^2$. Requiring $w'(r) = h(r)/\bigl(2 f(r)
  c^2\bigr)$, we can eliminate the off-diagonal term, and the prefactor of
  $\D r^2$ is simply renamed into $g(r)$. }

At large radii, gravitation will become negligible, so the metric
should approach Minkowskian form, hence we require 
\begin{align}
\lim_{r\to\infty} f(r) = 1\>, \qquad \lim_{r\to\infty} g(r) = 1\>.
\label{eq:bc_spherical}
\end{align}
So far, we have not used any physics, just symmetry. In order to
determine $f(r)$ and $g(r)$, we need to invoke physical ideas.

First, we make use of the equivalence principle.  Instead of
translating the physics in an accelerating system into terms of a
gravitating one, which requires to visualize two different but
equivalent systems in parallel, let us consider a freely falling
observer in the \emph{actual} system under consideration.\footnote{The
  first was Einstein's way of using the EP, whereas
  C.~Will\protect\cite{will95} is a strong proponent of the second
  way.}  The prescription then is to describe local physics in the
frame of that inertial observer by SR. For the freely falling
observer, there \emph{is} no gravitational field and everything that
the gravitational field does to CSOs must be due to the fact that they
are accelerating \bl{with respect to his inertial frame}.  Note that
an infinity of freely falling observers may be chosen at any point.
Normally, the best choice is to consider one that is momentarily at
rest with respect to the object (e.g., a CSO) that is to be described.

\begin{figure}[ht]
 \includegraphics*[width=7.0cm]{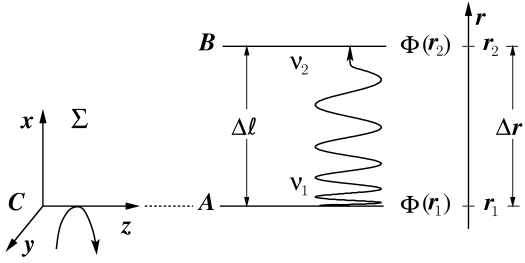}
 \caption{Equivalence principle: The inertial observer $C$ arrives at the
   apex of his trajectory next to $A$ the moment $A$ sends off a signal
   to $B$.}
 \label{fig:equival_princ}
\end{figure}

Using the EP, we obtain a relationship between the two functions to be
determined and the local gravitational acceleration. Consider two very
close CSOs $A$ at $r_1$ and $B$ at (the same angular position at)
$r_2$ with $r_2>r_1$, plus an inertial observer $C$ momentarily
comoving with $A$, the moment $A$ sends a light signal to $B$, as
depicted in Fig.~\ref{fig:equival_princ}. 

In $C$'s frame, the frequency $\nu_1$ of the signal is unchanged during its
short transit time $\Delta \tau$. But $B$, having accelerated to a
small velocity $v=a(r_2)\Delta \tau$, will receive it at a
reduced Doppler shifted frequency $\nu_2$. The deviation of the ratio
of frequencies from 1 is attributed to time dilation by $A$ and $B$,
who are stationary. From $\D s^2 = -c^2 \D\tau^2$, we read off that
the proper time of a CSO is given by $\D \tau = \sqrt{f(r)}\, \D
t$, so the frequency ratio may be calculated as
\begin{align} 
  \frac{\nu_1}{\nu_2} =
  \frac{\D\tau_2}{\D\tau_1}=\frac{\sqrt{f(r_2)}}{\sqrt{f(r_1)}}\>.
\label{eq:feq_ratio}
\end{align}
For $C$, the special
relativistic Doppler shift formula yields
\begin{align}
  \frac{\nu_1}{\nu_2} = \sqrt{\frac{c+v}{c-v}} \approx 1 + \frac vc \>.
\label{eq:doppler_spec}
\end{align}
Denote by $\Delta\ell$ the local proper distance
 separating $A$ and $B$. The metric is time orthogonal, therefore 
$\D\ell^2 = \D s^2\vert_{\D t=0}$. We then have, for the transit time,
 $\Delta\tau = \Delta\ell/c = \sqrt{g(r)}\, (r_2-r_1)/c$ (with
$r\in[r_1,r_2]$). 
Due to the closeness of the three observers and the smallness of their
relative motion, the length $\Delta\ell$ is the same for all of them during the
sequence of events considered.  It is again useful to introduce an
acceleration potential, describing local proper acceleration, via
$\D \Phi = a(r)\, \D\ell = a(r) \sqrt{g(r)} \>\D r$. This provides
\begin{align}
\frac{v}{c} \approx \frac{a \Delta\ell}{c^2} \approx \frac{\Delta \Phi}{c^2}\>,
\end{align}
and, for sufficiently small $\abs{r_2-r_1}$
\begin{align}
\sqrt{\frac{f(r_2)}{f(r_1)}} &= 1 + \frac{\Phi(r_2)-{\Phi(r_1)}}{c^2}\>,
\label{eq:time_dilation_fac_S}\\
a(r) &=  \frac{1}{\sqrt{g(r)}} \frac{\D \Phi}{\D r}\>.
\label{eq:proper_acc_S}
\end{align}
The first of these two equations is converted into a differential
equation as before
\begin{align}
\frac12 \frac{f'(r)}{f(r)} &= \frac{1}{c^2} \Phi'(r)
\label{eq:diffeq_f}
\end{align}
and this is solved by
\begin{align}
f(r) = \EXP{2\Phi/c^2}\>,
\label{eq:f_strong_field}
\end{align}
where the integration constant has been fixed by the boundary
condition \eqref{eq:bc_spherical}, given that, for large $r$, $\Phi$
must reduce to the Newtonian potential, i.e., go to zero in the
standard gauge. \bl{Since we do not know anything about $\Phi(r)$ for
  \emph{small} $r$, Eq.~\eqref{eq:f_strong_field} means no more in
  that $r$ range than expressing one unknown function, $f(r)$, in
  terms of another equally unknown one, $\Phi(r)$. All we achieve by
  this is to equate $f(r)$ to a quantity that has the physical
  interpretation of a potential. Nevertheless, the result is useful,
  as it is this interpretation that allows us to deduce the functional
  form of $f(r)$ at \emph{large} $r$ by the requirement that the potential
  become Newtonian there.}

To obtain a second relationship for the two functions $f(r)$ and
$g(r)$, let us look at the same thought experiment as in the case of a
uniform field. Assume that an observer at $r_0$ slowly lowers some
mass $m$ hanging from an inextensible massless tether towards smaller
$r$ values.

As before, we express the force using energy conservation,
\begin{align}
  F = -\frac{\D E(r)}{\D \ell} = -\frac{\D E(r)}{\D r}  \frac{\D r}{\D
    \ell}\>,
\end{align}
where $E(r)$ is the energy of the mass at position $r$, as judged by
the observer at $r_0$. By the same kind of argument as in the uniform
gravitational field we now obtain
\begin{align}
  E(r) &= \sqrt{\frac{f(r)}{f(r_0)}}\, m c^2
  \underset{r_0\to\infty}{\to} \sqrt{f(r)}\, m c^2\>,
\label{eq:energy_of_r}
\end{align}
where for simplicity our observer was moved to infinity. The
force felt at the upper end of the tether then is
\begin{align}
  F
& = -m c^2 \frac{\D \sqrt{f(r)}}{\D r} \frac{1}{\sqrt{g(r)}} =
  -m c^2 \frac{f'(r)}{2\sqrt{f(r) g(r)}}\>.
\label{eq:force_law}
\end{align}
Now we require that for large $r$, $\Phi(r)$ and $F(r)$ take
their Newtonian limits, i.e.,
\begin{align}
  \Phi(r) &= -\frac{G M}{r}\>, 
\label{eq:Phi_newton}\\
F(r) &= -\frac{G m M}{r^2}\>.
\label{eq:F_newton}
\end{align}
At this moment, we have no idea about what will become of
these laws as relativistic effects become strong, so we content ourselves
with determining a weak-field limit of the metric. The relevant
quantity distinguishing between weak and strong is $\Phi(r)/c^2$,
which outside the sun does not exceed $10^{-5}$ in our solar system,
so this limit should be appropriate for all calculations
referring to the latter.

Equations \eqref{eq:f_strong_field} and \eqref{eq:force_law} together
with expressions \eqref{eq:Phi_newton} and \eqref{eq:F_newton} for the
potential and force may be used to determine the two functions
$f(r)$ and $g(r)$:
\begin{align}
  f(r) &= \EXP{2\Phi/c^2} \approx 1+ 2\frac{\Phi}{c^2} = 1 - \frac{2G
    M}{r c^2}\>,
  \label{eq:result1_f_S}\\
   -\frac{G m M}{r^2} 
    &= -m
   \frac{G M}{r^2} \frac{1}{\sqrt{f(r)
      g(r)}}
  \nonumber\\
\Rightarrow\>\>   g(r) &= \frac{1}{f(r)} =
  \frac{1}{1-\frac{2G M}{r c^2}}\>.
\label{eq:result1_g_S}
\end{align}
This gives us, as a weak-field approximation, the line element
\begin{align}
  \D s^2 &= - \left(1 - \frac{2G M}{r c^2}\right)\,c^2 \D t^2 + \frac{1}{1-\frac{2G
      M}{r c^2}}\, \D r^2 
\nonumber\\
&\quad+ r^2\left(\D \vartheta^2 + \sin^2 \vartheta
    \,\D \varphi^2\right)\>,
\label{eq:metric_schwarzschild}
\end{align}
which is the exact result for the Schwarzschild metric!

Of course, this is too good to be true.
To see what happened, let us use, instead of the last formula from
\eqref{eq:force_law}, the first. This means, we 
approximate $\sqrt{f(r)}$ instead of $f(r)$ itself. Then the
calculation reads
\begin{align}
 \sqrt{f(r)} &= \EXP{\Phi/c^2} 
\approx 1+ \frac{\Phi}{c^2} = 1 - \frac{G M}{r c^2} \>,
 \nonumber\\
 -\frac{G m M}{r^2} &
 = -m \frac{G M}{r^2 }
  \frac{1}{\sqrt{g(r)}}
  \nonumber\\
\Rightarrow\>\>  g(r)&\equiv 1\>,
\end{align}
which is \emph{not} the expected result for $g(r)$.

In the two calculations, we used approximations for $f(r)$ which
agreed to first order in the small quantity $GM/r c^2$ ($\approx
\Phi/c^2$), but not to second order. However, the structure of the
equations is such 
that the \emph{first-order term of} $g(r)$ depends
on the \emph{second-order term of} $f(r)$.  This becomes immediately
clear, if we plug the approximation
\begin{align}
f(r) = 1 - \frac{2G M}{r c^2}+\beta_2 \left(\frac{G M}{r c^2}\right)^2
\end{align} 
into \eqref{eq:force_law} with the force law \eqref{eq:F_newton}. No
matter whether we take the formula with the derivative of $f(r)$ or
that with the derivative of its square root, we obtain the same
first-order result for $g(r)$, providing we expand the square root
correctly to second order:
\begin{align}
g(r) = 1 + 2\left(1-\beta_2\right) \frac{G M}{r c^2}\>.
\end{align}
Our first approximation corresponds to $\beta_2=0$, the second to $\beta_2=1$.

Therefore, to obtain a nontrivial result for $g(r)$,
we need to know $f(r)$ or, since the relationship
\eqref{eq:f_strong_field} between $f(r)$ and the potential is
exact,\footnote{The approximations used in applying the equivalence
  principle become exact in the limit of an infinitesimal distance
  between the two observers exchanging a light signal. Therefore, the
  differential equation \protect\eqref{eq:diffeq_f} is exact.} the
potential $\Phi(r)$ at least to the next order in ${G M}/{r c^2}$.
Knowing the exact result for $f(r)$ from GR, we may
infer that Eq.~\eqref{eq:Phi_newton} is indeed only a lowest-oder
approxima\-tion.\footnote{Equation \protect\eqref{eq:result1_f_S}
  contains \emph{two} approximations. The first is indicated by the
  $\approx$ sign, the second is the replacement of $\Phi(r)$ by its
  Newtonian limit. These two approximations happen to cancel each
  other to give the exact $f(r)$.  }

This immediately begs the question whether this is true for
\eqref{eq:F_newton} as well. Indeed, if we had to replace the force
law by $F(r) = - GmM/r^2 \left(1+\gamma_1 {G M}/{r
    c^2}+\ldots\right)$, this would bring in another unknown
coefficient and diminish our chances of calculating anything meaningful.

There are two reasons to believe -- without prior knowledge of the
exact result -- that \eqref{eq:F_newton} is, in fact, exact. The first
is that a similar thing happened in the case of the \emph{Rindler}
metric.  A force measured with tethers as
described turns out to be constant in space just as in a Newtonian
uniform gravitational field.  That the Newtonian force law be valid beyond the
weak-field approximation in the spherically symmetric situation as well was
one of the two postulates introduced by
Tangherlini.\cite{tangherlini62} It may be considered an extrapolation
from the homogeneous to the inhomogeneous case, hence a suggestive but
not really a strong argument. However, there is a much more convincing
way to justify this assumption.

The force $F(r)$ is a global field, measurable using
tethers.\footnote{An observer at finite $r_0$ will obtain a force
  field that has the same radial behaviour but is larger by a constant
  factor $1/\sqrt{f(r_0)}$.  Moreover, he will have to use poles
  instead of tethers to measure the field at $r>r_0$.} In principle,
the field may be measured in all of space (outside the central star).
Calculating the integral of the force over the surface of a sphere of
radius $r$, we obtain its total flux through that surface. If we
evaluate it on two concentric shells, the integral should not change,
if there is no source of the field between the two shells, i.e.~in
vacuum.  In fact, experience with both Newtonian gravity and
electrodynamics suggests this quantity to be a fixed multiple of the
total ``charge'' enclosed by the shells (mass, electrical charge) that
is at the origin of the field.  This charge should be a conserved
quantity -- in GR we expect it to be related to mass-energy.
Therefore, if such a conservation law prevails in GR, the force
\emph{must} be proportional to $1/r^2$ \emph{outside} the spherically
symmetric mass distribution, 
\emph{if} the area of the surface of a sphere grows as $r^2$. But we
defined our coordinate $r$ precisely so that this be the case.

Therefore, we will assume in the following that \eqref{eq:F_newton} is
exact.  This requirement goes beyond SR, the EP, and the Newtonian
limit. It is a partial replacement of Einstein's assumption that in
vacuum the Ricci tensor $R_{ik}$ must be required to
vanish.\cite{einstein16} In fact, it can be shown to follow from the
vanishing of the diagonal temporal component $R_{tt}$.  Being a much
weaker postulate than Einstein's, it gives us much less.  The latter
produces the exact Schwarzschild solution, the former fixes one
of the functions $f$ and $g$ in terms of the other. So if we knew the
expansion of $f(r)$ in powers of ${G M}/{r c^2}$, we could fully
calculate the corresponding expansion of $g(r)$. But at this stage, we
cannot even determine the coefficient $\beta_2$.

We conclude that just using the EP in trying to
transcend SR and Newtonian gravity gives us the first-order term of
the expansion of $f(r)$ in powers of ${G M}/{r c^2}$, but nothing more
($f(r)=1+\beta_1 {G M}/{r c^2}$ with $\beta_1=-2$), in accord with
Refs.~\onlinecite{schild60,gruber88}.  Our additional assumption about
the validity of \eqref{eq:F_newton} beyond the weak-field limit
produces a relationship between the coefficients of $f(r)$ and $g(r)$.
In particular, setting 
$ g(r) = 1/\left(1-\alpha_1 {G M}/{r c^2}-\ldots\right)$, we find
\begin{align}
\alpha_1 = 2 \left(1-\beta_2\right)\>.
\label{eq:rel_alpha1_beta2}
\end{align}

Essentially, our new postulate rests on the assumption that
mass-energy is the \emph{only} source of the gravitational field. It
is not expected to hold in alternative theories of gravity, in
which additional sources of the field are present. The Brans-Dicke
theory, for example, has a scalar field leading to a variable
effective gravitational constant. This bears some similarity to
electrodynamics in a medium with varying dielectric coefficient, in
which there would be apparent electrical charge distributions
leading to a non-vanishing divergence of the electric field
($\nabla\cdot\boldsymbol{E}\ne 0$). So the electrical field of a point charge
in such a polarizable medium would not fall off as $1/r^2$. Indeed,
checking whether the postulate is satisfied in spherically symmetric
solutions of the Brans-Dicke theory, we find that it is not, unless
the scalar field is constant. 

To obtain the exact result for $f(r)$ and $g(r)$, we would need a
second postulate. While it is possible to generate a more plausible
postulate than Tangherlini's second one, and while this postulate is not
subject to Rindler's criticism,\cite{rindler69} both its physical
justification and its practical use are somewhat more complex than
that of the postulate about mass-energy conservation invoked so far
(albeit still simpler than the field equations). Therefore, the
presentation of that route to the exact Schwarzschild metric will be
postponed to a different publication. It may not really be suited for
the classroom at the early stage envisaged.
Instead, we will take the point of view here, that if we cannot determine
either $\alpha_1$ or $\beta_2$ from \emph{theoretical} arguments, why
not turn to \emph{experiments}?

\section{The perihelion precession of Mercury}
\label{sec:perihelion_prec}
As it is assumed that the class has no prior knowledge on GR, we first
need to provide an approach to the equations of motion of a particle
in a given metric, i.e., the geodesic equations. This is done in the
argument from Eqs.~\eqref{eq:lagrangian_schwarzschild} to
\eqref{eq:geodesic_theta}, based on the EP. If the material of this
paper is used at a later stage in a GR course and the geodesic equations are
already known, this argument may be skipped and reference can be made
to the standard approach to the geodesic equations via the exact form
of the Lagrangian given in \eqref{eq:lagrangian_schwarzschild}.

We write the line element as
\begin{align}
  \D s^2 &= -c^2 \D\tau^2 = -\EXP{2\Phi(r)/c^2}\,c^2 \D t^2 +
  \frac{1}{1-\alpha(r)}\, \D r^2
  \nonumber\\
  &\quad + r^2\left(\D \vartheta^2 + \sin^2 \vartheta \,\D
    \varphi^2\right)
\label{eq:line_el_prec}
\end{align}
with
\begin{align}
  \EXP{2\Phi(r)/c^2} &= 1 - \frac{2\tilde M}{r}+\beta(r)\>, \qquad
  \tilde M = \frac{GM}{c^2}\>,
  \\
  \beta(r) &= \beta_2 \frac{\tilde M^2}{r^2} +
  \mathcal{O}\left(\frac{\tilde M^3}{r^3}\right) \>,
  \\
  \alpha(r) &= \alpha_1 \frac{\tilde M}{r} + \alpha_2
  \frac{\tilde M^2}{r^2} + \mathcal{O}\left(\frac{\tilde
      M^3}{r^3}\right)\>.
\label{eq:metric_expansion}
\end{align}
$M$ is assumed to be the mass of the sun and we treat a planet
(Mercury) with mass $m$ moving in its gravitational field. Consider now
the quantity (an overdot signifies a derivative with respect to proper
time)
\begin{align}
  L &= \frac m2 \left(\frac{\D s}{\D\tau}\right)^2 = \frac m2
  \left(\rule{-1mm}{5mm}\right.\frac{1}{1-\alpha(r)} \dot{r}^2
  \nonumber \\
  &\quad+ r^2 \dot\vartheta^2 + r^2 \sin^2 \vartheta
  \,\dot{\varphi}^2\left.\rule{-1mm}{5mm}\right) - \frac m2
  \EXP{2\Phi(r)/c^2} c^2 \dot{t}^2 \nonumber \\ &\approx \frac m2
  \left(\dot{r}^2 + r^2 \dot\vartheta^2 + r^2 \sin^2 \vartheta
    \,\dot{\varphi}^2\right) - \frac m2 c^2 \,\dot{t}^2 - m \Phi(r)
  \,\dot{t}^2
  \nonumber \\
  &\approx \frac m2 \left(r_t^2 + r^2 \vartheta_t^2 + r^2 \sin^2
    \vartheta \,\varphi_t^2\right) - \frac m2 c^2 - m \Phi(r)
  \nonumber \\
  & = T-V - \frac m2 c^2\>.
\label{eq:lagrangian_schwarzschild}
\end{align}
Herein, the first approximation uses the smallness of $\tilde M/r$ and
the second takes advantage of the fact that a planet moves slowly in
comparison with the speed of light, so $\dot t\approx 1$, i.e., global
and proper time are almost the same, and derivatives with respect to the
proper time may be replaced by derivatives with respect to $t$. The
last line finally identifies the kinetic energy $T$ and the potential
energy $V$ in the Newtonian limit. The NL of $L$ is a Newtonian
Lagrangian, which suggests that $L$ itself might be a relativistic
Lagrangian.

This conjecture can in fact be proven on the basis of the EP.  What
the principle tells us is that in a freely falling frame the local
laws of motion are those of SR. Hence, in such a frame, obtainable by an
appropriate local coordinate transformation from the global metric, a
point mass moves along a straight line, which we can determine from
local initial conditions, then transform back to obtain a piece of the
trajectory in the global frame, which gives the initial conditions for
the next (close-by) local frame to which we may transform. Continuing
the procedure, we obtain a piecewise construction of the trajectory. A
more elegant way rather than to construct pieces of the trajectory
is to produce equations of motion in the global frame from those of the
local frames and then find the full solution in the global frame
directly.  Now it is obvious that with the Minkowski line element the
quantity
\begin{align}
  L' = \frac{m}{2}\left(\frac{\D s}{\D\tau}\right)^2 =
  \frac{m}{2}\left(\dot{X}^2+\dot{Y}^2+\dot{Z}^2-c^2 \dot{T}^2\right)
\label{eq:lagrangian_minkowski}
\end{align}
is a valid Lagrangian for special relativistic motion of a free
particle. All coordinates are cyclic, so the equations of motion state
that $T$, $X$, $Y$, $Z$ are linear functions of the proper time, which
means that the four-velocity is constant. These are the correct
equations of motion in SR. Transforming this Lagrangian back to
the global frame is trivial, because both $\D s$ and $\D \tau$ are
relativistic invariants (as is $m$), so the result of the
transformation is $L$ of Eq.~\eqref{eq:lagrangian_schwarzschild}.
Hence, we can derive the equations of motion in the metric from
$L$.

Are there solutions with $\vartheta =
\pi/2=\text{const.}$  as in the Newtonian case? The equation of
motion for $\vartheta$
\begin{align}
  \frac{\D}{\D\tau} \frac{\partial L}{\partial \dot\vartheta} -
  \frac{\partial L}{\partial \vartheta} = \frac{\D}{\D\tau} m r^2
  \dot\vartheta - m r^2 \sin\vartheta \cos\vartheta \,\dot\varphi^2 = 0
\label{eq:geodesic_theta}
\end{align}
is obviously solved by $\vartheta \equiv \frac{\pi}{2}$, so we may
indeed restrict ourselves to motion in the equatorial plane.

Both $\varphi$ and $t$ are cyclic coordinates leading to conservation laws:
\begin{align}
  r^2 \dot\varphi &= h = \text{const.}
\label{eq:angular_momentum_cons}
\\
\EXP{2\Phi/c^2} \dot t &= k = \text{const.}
 \label{eq:energy_cons}
\end{align}
Equation \eqref{eq:angular_momentum_cons} describes conservation of
the component of angular momentum perpendicular to the equatorial
plane, and Eq.~\eqref{eq:energy_cons} expresses energy conservation.

Finally, instead of writing down the Euler-Lagrange equation for $r$
(the many $r$ dependent terms would lead to a messy formula), we
exploit the constancy of the Lagrangian itself\footnote{That the
  Lagrangian is constant here is a consequence of its definition in
  terms of invariants.  We have $\D s^2 = -c^2 \,\D\tau^2$, hence
  $L=-m c^2/2$.  On the other hand, the constancy of the
  \emph{Hamiltonian} in classical mechanics follows from a
  conservation law, energy conservation, implied by invariance under
  time translations. But the Hamiltonian itself is not an invariant
  under arbitrary coordinate tranfsormations. In classical mechanics,
  the Lagrangian of a relativistic free particle is $L=-m
  c^2\sqrt{1-v^2/c^2}$, which is \emph{not} a constant. The difference
  comes from the fact that there the action integral is $\int L \D t$,
  where $t$ is the time of some inertial system, whereas here we
  define it as $\int L \D \tau$, and the relationship between proper
  time $\tau$ and $t$ is, of course, $\D\tau=\sqrt{1-v^2/c^2}\,\D t$.}
to obtain another integral of the motion:
\begin{align}
  -c^2 = - \EXP{2\Phi/c^2} c^2 \dot t^2 + \frac{1}{1-\alpha(r)}
  \dot{r}^2 + r^2 \dot\varphi^2\>.
\label{eq:r_equation}
\end{align}
Using \eqref{eq:angular_momentum_cons} and \eqref{eq:energy_cons}, we
can separate out an equation for the radial coordinate alone
\begin{align}
  \frac{1}{1-\alpha(r)} \dot{r}^2 + \frac{h^2}{r^2} + \left(1-k^2
    \EXP{-2\Phi/c^2}\right) c^2 = 0\>.
\label{eq:r_equation_1}
\end{align}
(The Newtonian limit of this equation is obtained by letting
$c\to\infty$, which implies $k\to 1$ and leads to the familiar $\dot
r^2 + h^2/r^2 - 2GM/r = 0$.)  We are interested in the spatial
trajectory only, i.e., the function $r(\varphi)$, so we write $\dot r
= \D r/\D\varphi \>\D\varphi/\D\tau = r' \dot\varphi = r' h/r^2$. It
is then convenient to introduce the new variable $u(\varphi) = 1/r$,
whence $u' = - r'/r^2$. 
Using the expansions of $\alpha(r)$ and $\exp(2\Phi(r)/c^2)$, 
multiplying the equation by $[1-\alpha(r(u))]/h^2$ and expanding all
terms to second order in $ \tilde M u$, we obtain after a
rearrangement of terms:
\begin{align}
  u'^2 &+ u^2 - \alpha_1 \tilde M u^3 - \alpha_2 \tilde M^2 u^4
  \nonumber\\
  & -\frac{c^2}{h^2}\left(2 \tilde M u +k^2 (4-2\alpha_1-\beta_2) 
    \tilde M^2 u^2\right)
  \nonumber\\
  & + \frac{c^2}{h^2}\left(1-k^2\right)\left(1 + (2-\alpha_1)
    \tilde M u - \alpha_2 \tilde M^2 u^2\right) = 0\>.
\label{eq:r_equation_3}
\end{align}
In order to simplify this equation, we consider the sizes of its
terms. The leading order terms are $u'^2 + u^2 - 2c^2 \tilde M u/h^2 +
c^2 (1-k^2)/h^2$. While $\tilde M u$ is very small,\footnote{An
  estimate for $\tilde M u$ is provided by $G M/c^2 a\approx
  2.55\times10^{-8}$, where $a\approx 57.9\times10^9\>$m is the
  semi-major axis of Mercury's orbit.} the first term linear in $u$ is
multiplied by $c^2$, a large factor. This is the reason why we have to
take into account the $u^2$ term in the first parentheses. However, we
may drop the $\alpha_2$ term in the first line and also the one
multiplied by $c^2 (1-k^2)/h^2$, because in this term the large factor
$c^2$ is compensated by the small factor $1-k^2$. To see this, let us
estimate $k$, by evaluating Eq.~\eqref{eq:r_equation_1} at the
perihelion, where $\dot r=0$, and taking the NL. The Newtonian value
for the minimum distance of the planet to the center of motion is
$r_{\text{min}} = h^2/(GM(1+e))$, where $e$ is the eccentricity and
$h$ twice the areal velocity, referred to Newtonian time instead of
proper time. We find
\begin{align}
  k^2 - 1 & = -\frac{GM}{c^2 a}\>,
\label{eq:k_determin}
\end{align}
where $a=r_{\text{min}}/(1-e)$ is the semi-major axis of the orbital
ellipse of the planet considered.  The result is twice the orbital
energy of the  planet divided by $m c^2$, a very small quantity
indeed.

Having justified the neglect of the $\alpha_2$ terms in
\eqref{eq:r_equation_3}, we take the derivative with respect to
$\varphi$ (to obtain a linear lowest-order equation), and get, after
dividing off the common factor $2u'$
\begin{align}
  u'' +& u = \frac{GM}{h^2} + \tilde M
  \left(\rule{0mm}{5mm}\right.\!\frac32 \alpha_1 u^2 +
  k^2\left(4-2\alpha_1-\beta_2\right)\frac{GM}{h^2} u
  \nonumber\\
  &\qquad-\frac{c^2 }{h^2}
  (1-k^2)\left(1-\frac{\alpha_1}{2}\right)\!\left.\rule{0mm}{5mm}\right)\>.
\label{eq:perturb_lin_eq}
\end{align}
Herein, we may consider the term multiplied by $\tilde M$ a
small perturbation, first solve the equation with $\tilde M$ set
equal to zero and then correct the result using perturbation theory.
The lowest-order equation is $u''+u= GM/h^2$ and it is solved by
\begin{align}
  u_0 = \frac1{r_0} = \frac{GM}{h^2} \left(1+e \cos\phi\right)\>,
\label{eq:sol_newton}
\end{align}
the well-known Newtonian result. The eccentricity $e$ is one of the
integration constants. Another would be the angle $\varphi_0$ between
the semi-major axis and the $x$ axis, which has been absorbed into a
redefinition of the $x$ axis, so the perihelion is at $\varphi=0$. The
results for $r_{\text{min}}$ and $a$ used in deriving \eqref{eq:k_determin}
immediately follow from \eqref{eq:sol_newton}. Note that only the term
$\propto u^2$ has to be treated within perturbation theory. Without
it, Eq.~\eqref{eq:perturb_lin_eq} would be solvable exactly. Because
we still would have to treat one term perturbatively, we might as well
consider all terms multiplied by $\tilde M$ perturbations.

Now we iterate the equation, inserting $u_0$ on the right-hand side,
to obtain the first-order correction:
\begin{align}
  u'' +& u = \frac{GM}{h^2}
  + \tilde M \frac{G^2M^2}{h^4} \left[\rule{0mm}{5mm}\right.\!\frac32
  \alpha_1 \left(\rule{0mm}{4.5mm}\right.\!1+\frac{e^2}{2} + 2 e \cos\varphi
 \nonumber\\
 & + \frac{e^2}{2}
    \cos 2\varphi\left.\rule{0mm}{4.5mm}\!\right)
  +k^2\left(4-2\alpha_1-\beta_2\right) \left(1+e
    \cos\varphi\right)\!\left.\rule{0mm}{5mm}\right]
  \nonumber\\
  &-\tilde M \frac{c^2}{h^2}(1-k^2)\left(1-\frac{\alpha_1}{2}\right)\>.
\label{eq:secular}
\end{align}
This is the equation of motion of a driven harmonic oscillator with
resonant terms on the right-hand side (the terms
$\propto\cos\varphi$). A straightforward treatment would lead to
self-amplifying solutions, destroying the applicability of
perturbation theory. Therefore, we use a slightly more sophisticated
approach, the Poincaré-Lind\-stedt method, in which the argument of
the solution is considered a function of the perturbation, too.  
Restricting ourselves to the lowest-order scheme, we write
$u(\varphi)=\tilde u((1+\varepsilon)\varphi)$, with $\varepsilon$
being proportional to the small parameter $\tilde M$. We then have
$u''+u = (1+\varepsilon)^2 \tilde u'' + \tilde u \approx
(1+2\varepsilon) \tilde u'' + \tilde u$ and setting $\tilde u = \tilde
u_0 + \tilde M \tilde u_1$, we obtain
$\tilde u''_0 + 2\varepsilon\tilde u''_0 + \tilde M \tilde u''_1 +
\tilde u_0 + \tilde M \tilde u_1$
on the left-hand side of \eqref{eq:secular}. Since $\tilde u_0 = u_0$,
the term multiplied by $\varepsilon$ is proportional to $\cos\varphi$
and by an appropriate choice of $\varepsilon$, we may cancel the
secular terms on the right-hand side. Then $\tilde u_1$
satisfies an equation of the type $ \tilde u''_1+ \tilde u_1 =
A+B\cos2\varphi$, but we are not particularly interested in solving
it, as the information about the perihelion precession is in the
modified periodicity of the solution, determined already by the value of
$\varepsilon$.  The new period is $P=2\pi/(1+\varepsilon) \approx
2\pi(1-\varepsilon)$, hence the perihelion shift $\Delta P$ per period
is
\begin{align}
  \Delta P = -2\pi\varepsilon = \pi \tilde M \frac{GM}{h^2}
  \left(\alpha_1 -\beta_2+4\right)\>,
\end{align}
where we have replaced $k^2$ by 1.  The formula may be recast in terms
of more convenient quantities. $h$ is twice the areal velocity, hence
in the approximation of a Kepler ellipse
\begin{align}
  h=2 \frac{\pi a b}{T} = 2 \frac{\pi a^2\sqrt{1-e^2}}{T}\>,
\end{align}
with $T$ its orbital period, and from Kepler's third law
\begin{align}
T^2 = \frac{4\pi^2 a^3}{GM}\>,
\end{align}
we obtain an expression for $GM$. Combining the two results, we have
$(GM)^2/h^2 = 4\pi^2 a^2/(T^2(1-e^2))$ and find
\begin{align}
  \Delta P = \frac{ 4\pi^3 a^2}{c^2
    T^2(1-e^2)}\left(\alpha_1-\beta_2+4\right)\>.
  \label{eq:perihelion_shift_per_orbital}
\end{align}
The \emph{annual} perihelion shift $\Delta P_a$ is obtained from this by
multiplying with $T_{\text{earth}}/T$, which is a factor of 4.152 for Mercury, having an
orbital period of 87.969 d. Plugging in numbers, we get
$\Delta P_a =  3.4730\times10^{-7} \left(\alpha_1-\beta_2+4\right)$.
This is the result in radians. To convert it to arcseconds, we note
that $1'' = 2\pi/360/3600\> \text{rad}= 4.8481\times10^{-6}\>
\text{rad}$.  Then we have $\Delta P_a = 0.0716''\times
\left(\alpha_1-\beta_2+4\right)$.  Experimental measurements give
$\Delta P_a=0.4298''$,\cite{wiki_tests_gen_rel} from which we infer
\begin{align}
\alpha_1-\beta_2+4 \approx \frac{0.4298''}{0.0716''} = 6.003\>.
\end{align}
This immediately leads to the conjecture
\begin{align}
\alpha_1-\beta_2 = 2\>.
\end{align}
Together with $\alpha_1=2(1-\beta_2)$ from
Sec.~\ref{sec:schwarzschild}, we end up with
\begin{align}
  \alpha_1 &= 2\>,\quad
\beta_2 = 0\>.
\end{align}
Therefore, we have now inferred the Schwarzschild metric
\eqref{eq:metric_schwarzschild} to second-order accuracy in the small
parameter $GM/rc^2$ for the coefficient $g_{tt}=f(r)$ and to
first-order accuracy for $g_{rr}=g(r)$, which is known as the first
parameterized post-Newtonian approximation (PPN).\cite{misner73}
Clearly, the two parameters determined in this section from
experimental information would be known only with finite precision.
They could not be claimed to be exact without the benefits of the
field theory.

We conclude that Einstein might indeed have used a similar approach in
1911 and would then have been able to correctly predict gravitational
light deflection by the sun five years earlier than he actually did.
Also, he might have found the Schwarzschild solution before
Schwarzschild\cite{schwarzschild16a} and Droste.\cite{droste17}

As it turns out, experimental information on light deflection gives
much simpler access to the coefficient $\alpha_1$ than peri\-helion
precession data.  So let us turn to a brief analysis of the behavior
of light in the gravitational field of a spherically symmetric mass
distribution.

Unfortunately, this experimental informaton was not available before
1919 and then only with low accuracy.\cite{dyson20}

\section{Light deflection}
\label{sec:light_bending}
Again, if the geodesic equations are supposed known, part of this
section can be skipped. Even then, the use of isotropic Schwarzschild
coordinates suggested here is favorable, as it leads to simpler
equations and allows one to repeat the argument about the factor of 2
between the EP prediction and the full calculation, given below.
Given the equations for null geodesics, one may eliminate the affine
parameter, determine the energy constant $k$ from the limit
$r\to\infty$ and introduce the variable $u=b/\rho$, which produces
Eq.~\eqref{eq:energy_harmonic_osc}.

Assuming instead no prior knowledge of the geodesic equations, it may be
argued that just as the principle of least action governs the motion
of particles and gives us the equations of motion, once we know the
Lagrangian, Fermat's principle governs the paths of light and gives us
their equations, once we know the index of refraction or,
equivalently, the speed of light. Both principles have the advantage
of being coordinate free, so we may expect them to work in curved
spacetime without problems. It should not matter whether a given
coordinate system describes a patch of flat or of curved spacetime.

Let us therefore put Fermat's principle to use in
describing light deflection. It is obvious that the coordinate
speed of light, obtained from the line element by setting $\D
s^2=0$, varies in a metric such as \eqref{eq:spherical_static}. We may
interpret this in terms of a variable refractive index $n$,
and then require, in order to calculate the path of light
\begin{align}
  \delta S_F = 0\quad \text{for}\quad S_F= \frac1c \int \!\D l\, n(l)
  = \!\int \!\frac{\D l}{c(l)}\>,
\label{eq:fermat}
\end{align}
where $\D l$ is the (coordinate) length element of the 
path of the light ray and $c(l)$ is the local speed of light. The
endpoints of the path are supposed to be fixed. Hereafter, I
will slightly abuse notation in taking $c$ for the universal speed of
light and $c(x)$ for the coordinate speed of light at
some point $x$. 

When written in the form \eqref{eq:fermat}, the principle requires
$n(x)$ and hence $c(x)$ to be a scalar function, i.e., the velocity of
light should be isotropic. This is clearly not true for the general
metric \eqref{eq:spherical_static}. The velocity of light in the
radial direction is given by $c_r(r) = {\D r}/{\D t} = {\sqrt{f(r)}\,
  c}/{\sqrt{g(r)}}$ (setting $\D s$, $\D\vartheta$ and $\D\varphi$
equal to zero), whereas the transverse speed of light is $c_t(r) =
\sqrt{f(r)}\, c$ (setting $\D s$ and $\D r$ equal to zero).

\begin{figure}[ht]
  \includegraphics*[width=8.0cm]{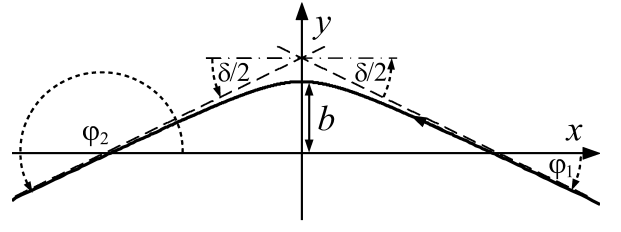}
  \caption{Pictorial representation of the trajectory of a photon
    and its orientation in the coordinate system
    with $x=\rho\cos\varphi$, $y=\rho\sin\varphi$. Small angles are
    generously exaggerated in size.}
\label{fig:hyperbola}
\end{figure}

While it is possible to phrase Fermat's principle for an\-isotropic
light propagation,\cite{weyl17,kovner90,nityananda92} the necessity to
first derive this unfamiliar formulation makes it unattractive for
work in class. Instead, we rewrite our metric to spatially isotropic
form. This can be achieved via introduction of a new radial coordinate
$\rho$ as follows: set $r=r(\rho)$ and require
\begin{align}
g(r) \,r'(\rho)^2 = \frac{r^2}{\rho^2}\>. 
\end{align}
Solving this differential equation for $r$ gives a metric with spatial
part $g(r(\rho))\> r'(\rho)^2\left(\D\rho^2 + \rho^2 \D\vartheta^2 +
  \rho^2 \sin^2\vartheta\,\D\varphi^2\right)$. For the metric
\eqref{eq:line_el_prec} with $\alpha(r)$ approximated by the first term of
\eqref{eq:metric_expansion}, this procedure leads to
\begin{align}
r = \rho\left(1+\frac{\alpha_1\tilde M}{4\rho}\right)^2\>,
\end{align}
where the integration constant has been chosen so that far from the
central mass both coordinates become equal to each other.  In the new
coordinates, the spacetime line element takes the form
\begin{align}
\D s^2 &= -k_1^2 \,c^2\,\D t^2 + k_2^2 \, \D l^2 \>,
\\
\shortintertext{where}
\D l^2 & = \D\rho^2 + \rho^2 \D\vartheta^2 +
  \rho^2 \sin^2\vartheta\,\D\varphi^2\>,
\nonumber\\
k_1 &= \left(1-\frac{2\tilde M}{\rho}\right)^{1/2}
 + \mathcal{O}\left(\frac{\tilde M^2}{\rho^2}\right)\>,
\nonumber\\
k_2 &= \left(1+\frac{\alpha_1\tilde M}{4\rho}\right)^2
+ \mathcal{O}\left(\frac{\tilde M^2}{\rho^2}\right)\>.
\end{align}
The local coordinate speed of light is then given by 
\begin{align}
  c(\rho) = \frac{k_1}{k_2}\, c = \left(1-\frac{(2+\alpha_1)\tilde M}{2\rho}\right) c +
  \mathcal{O}\left(\frac{\tilde M^2}{\rho^2}\right)\>.
\label{eq:local_light_sp}
\end{align}
For symmetry reasons, we expect the path of a light ray in the
equatorial plane $\vartheta =\pi/2$ to remain in that plane, so we can
drop the $\D\vartheta$ contribution to the spatial coordinate line
element $\D l$. We are then left with the task to minimize
\begin{align}
\hspace*{-2mm}S_F = \!\int \frac{\sqrt{\D\rho^2+\rho^2\,\D\varphi^2}}{c(\rho)} 
= \!\int \frac{\sqrt{\rho'(\varphi)^2+\rho(\varphi)^2}}{c(\rho(\varphi))} \,\D\varphi\>.
\end{align}
Taking the integrand to be a function
$s_F(\rho'(\varphi),\rho(\varphi),\varphi)$, we note it does not
depend on $\varphi$ explicitly, so the ``Hamiltonian'' 
\begin{align}
  \rho'\frac{\partial s_F}{\partial\rho'} - s_F
  \end{align}
is constant.
Naming the constant $-b/c$, we find
\begin{align}
  \frac{\rho^2}{\sqrt{{\rho'}^2+\rho^2}} = b \,\frac{c(\rho)}{c}\>.
\label{eq:basic_diffeq}
\end{align}
Solving this algebraically for $\rho'$ and substituting $u=b/\rho$, we obtain
\begin{align} 
{u'}^2 + u^2 = \frac{c^2}{c(\rho(u))^2} = 1
  +\frac{(2+\alpha_1)\tilde M u}{b} + \mathcal{O}\left(\frac{\tilde M^2}{b^2}\right)\>.
\label{eq:energy_harmonic_osc}
\end{align}
Differentiating with respect to $\varphi$ and dividing through by
$2u'$, we end up with an extremely simple equation
\begin{align}
u'' + u = \frac{(2+\alpha_1)\tilde M}{2b}\>.
\label{eq:harmonic_osc}
\end{align}
This is a shifted harmonic oscillator. Rewritten in terms of $\rho$, the solution 
reads
\begin{align}
  \frac{b}{\rho} &= \sin\left(\varphi-\varphi_0\right) + \frac{(2+\alpha_1)\tilde
      M}{2b} \>.
\label{eq:light_hyperbola}
\end{align}
From this equation, describing a
hyperbola, we can read off the deflection angle.  For convenience, we set
$\varphi_0=0$, which gives the hyperbola the orientation shown in
Fig.~\ref{fig:hyperbola} (assuming $b>0$).

For $\rho\to\infty$, given the smallness of the second term on the
right-hand side of Eq.~\eqref{eq:light_hyperbola}, the argument of the
sine must go to $\varphi_1=-(1+\alpha_1/2)\tilde M/b$ for $x>0$ and to
$\varphi_2=\pi+(1+\alpha_1/2)\tilde M/b$ for $x<0$.  The deflection
angle then is
\begin{align}
  \delta \equiv \varphi_2 -\pi-\varphi_1 =
  \left(1+\frac{\alpha_1}{2}\right)\frac{2GM}{b c^2}\>.
\end{align}
With $\alpha_1=2$, Einstein's 1916 result $\delta = 4GM/(b c^2)$ is
recovered, corresponding to an angle of 1.75'' for a light ray that
grazes the surface of the sun (i.e., when $b$ is equal to the radius of the
sun).

The utility of isotropic coordinates in the calculation of this effect
may be underlined by a comment regarding the factor of 2
between the correct (first-order) result and Einstein's original result. In his 1911
paper, Einstein derived the speed of light from the equivalence
principle alone, which means that he replaced the local gravitational
field by a patch of an inertial field 
with a value $g_{tt}$ adapted to the true
field, but effectively with $g_{rr}=1$. He therefore obtained 
\begin{align}
c(r) = \left(1+\frac{\Phi(r)}{c^2}\right) c = \left(1-\frac{\tilde M}{r}\right) c\>.
\end{align}
Plugging this velocity into Fermat's principle, we get a formula in
which the deviation of the coordinate speed of light from the vacuum
speed of light is formally half the value of the deviation in the
isotropic Schwarzschild metric, if we rename $r$ to $\rho$ (compare
with Eq.~\eqref{eq:local_light_sp} for $\alpha_1=2$). Since the
angular deflection is so small that it is easily captured by
perturbation theory, the final result must be \emph{linear} in this
deviation (from the zeroth-order straight-line solution), so the angle
of deflection obtained in the full theory must be a factor of 2 larger
than the one obtained by the EP only. No argument of comparable
simplicity is available when comparing a calculation based on the EP
with the full theory using the \emph{original} Schwarzschild metric.
In that metric, the speed of light is anisotropic, agreeing with the
prediction from the EP for \emph{transverse} light rays and differing
from it for \emph{radial} ones. (Far from the sun, the ray is
essentially radial.)

Note that if we assume the \emph{experimental result} on light deflection
to be available, we can deduce the value of $\alpha_1$ without the
need to require that \eqref{eq:F_newton} is valid beyond the lowest
order in $\tilde M$. Hence the experiment allows us to determine this
coefficient without any second-order knowledge of $f(r)$, whereas
we need this information, if we want to obtain $\alpha_1$ from the
perihelion precession. How does this come about?

Consider the complete Lagrangian of
Eq.~\eqref{eq:lagrangian_schwarzschild}. In it, the term $\propto \dot
r^2$ is much smaller than the term $\propto c^2 \dot t^2$, because the
velocity $\dot r$ of a planet is much smaller than $c$. So in order to
obtain an accurate estimate of the small quantity $\alpha_1 \tilde
M/r$ appearing as a factor in the former term, we have to know much
smaller factors of the latter term, i.e., we have to calculate $f(r)$
to second-order precision. The case of light bending is different.
Here the $\D r^2$ and $\D t^2$ terms of the line element are the same
order of magnitude, because for light $\D r/\D t$ is on the order of
$c$.  Thus, to determine $g_{rr}$ accurate to first order from
experiment, it is sufficient to know $g_{tt}$ to first order.

\section{Conclusions}
\label{sec:conclusions}

It is well-known that out of the three classical tests of general
relativity, the gravitational redshift, essentially explicable in
terms of the EP, does not probe the field equations,
whereas light deflection in a gravitational field and the perihelion
precession do depend on them. The same is true for the fourth test
conceived later, the Shapiro delay, not discussed in
detail here.
 
We have explored in this paper, how far some simple ideas, not
exploiting the full field theory, may carry us in determining a
usable weak-field approximation of the metric outside a spherically
symmetric mass distribution. This amounts to the approximate
construction of no more than two radial functions. As it was not
expected that these ideas would generate enough information to predict
spacetime curvature quantitatively, we were willing to accept one
adjustable parameter to emerge from one of the three experiments
probing the field equations, which would then, hopefully, allow us to
make quantitative predictions of the other two. In part, this was
motivated by the increase in credibility that a ``physics first''
approach to GR would gain, if the metric employed to derive
predictions could be justified without use of the field equations.

It turned out that this program is feasible but that the two
experiments considered need different levels of additional information.

In the case of the perihelion precession, to progress at all we had to
make an assumption about the range of validity of the force law
\eqref{eq:F_newton}, because \emph{two} unknown coefficients of the
metric are needed in a lowest-order PPN description. Given that new
assumption and a measurement of the perihelion precession, we get the
metric with sufficient accuracy to predict both the outcome of the
light bending experiment and the Shapiro effect
quantitatively.\footnote{I have not discussed the Shapiro delay here,
  but it is clear that its quantitative description is possible as
  soon as we have a metric that accurately describes gravitational
  light deflection.}

In the case of light bending by the sun, only one parameter is missing
in the metric as far as it is determined by the EP and the NL.  A
measurement of light deflection fixes this parameter.  With the metric
so obtained, the Shapiro delay could be predicted quantitatively
without any additional postulate on the force law.  However, the same
metric would be insufficient to quantitatively predict the perihelion
precession of Mercury. Assuming the force law to be accurate beyond
lowest order, enough information can be gathered.

In regard to research, our results may be considered not overly
interesting, since the exact solution of the field equations is known.
But they might be of some use in the classroom. One of the
calculations from Secs.~\ref{sec:perihelion_prec} and
\ref{sec:light_bending} can be discussed during a course, the other
may be assigned as a homework. Physical understanding of the
spherically symmetric system would be greatly improved. The
experience that our attempt at a simplification, avoiding the field
equations, involves moderately extensive calculations at least in the
case of the perihelion precession, will help students to appreciate
the introduction of the field equations later, which allow, with a little 
more effort, to obtain the \emph{exact} form of the spherically
symmetric static metric.

\textbf{Acknowledgment} I am grateful to Robert Shuler from NASA,
Johnson Space Center, who set me on track and largely inspired this
work. His idea was to start from the result for the measurement of
forces with a tether in the Schwarzschild metric and to invert the
procedure in order to obtain the metric from the force law. When this
did not quite work out, he suggested to use experimental results to
get the missing information.
\rule{0pt}{0pt}\\[-4mm]
\newcommand{\phre}[1]{Phys. Rev. E {\bf #1}}
\newcommand{\phrl}[1]{Phys. Rev. Lett. {\bf #1}}
\bibliographystyle{apsrev4-1long}
%

\end{document}